\documentclass[preprint]{elsarticle}

\usepackage{lineno,epstopdf,amsmath,longtable,siunitx,subcaption,graphics}
\modulolinenumbers[5]
\begin{document}

\begin{frontmatter}

\title{Presence-absence estimation in audio recordings of tropical frog communities}%\tnoteref{mytitlenote}}





%% Authors
\author[puceit]{Andr\'es Estrella~Terneux}
\ead{eestrella016@puce.edu.ec}
\author[puceit]{Dami\'an~Nicolalde\corref{cor1}}
\ead{danicolalde@puce.edu.ec}
\author[puceit]{Daniel~Nicolalde}
\ead{danielnicolalde@gmail.com}
\author[puce]{Andr\'es Merino-Viteri}
\ead{armerino@puce.edu.ec}
%\author[puceit]{Gustavo~Chafla}
%\ead{gxchafla@puce.edu.ec}

\cortext[cor1]{Corresponding author at: Laboratorio de Tecnolog\'ias de la Informaci\'on LTIC, Facultad de Ingenier\'ia, Pontificia Universidad Cat\'olica del Ecuador, P.O. Box: 17 01 21 84, Av. 12 de Octubre 1076 y Roca, Quito, Ecuador, Tel.:+593 22991700. }
%\cortext[cor2]{Principal corresponding author}
%\fntext[fn1]{This is the specimen author footnote.}
%\fntext[fn2]{Another author footnote, but a little more
%	longer.}
%\fntext[fn3]{Yet another author footnote. Indeed, you can have
%	any number of author footnotes.}
\address[puceit]{Laboratorio de Tecnolog\'ias de la Informaci\'on, Facultad de Ingenier\'ia, Pontificia Universidad Cat\'olica del Ecuador,  Av. 12 de Octubre 1076 y Roca, Quito, Ecuador}
\address[puce]{Laboratorio de Ecofisiolog\'ia, Escuela de Ciencias Biol\'ogicas, Pontificia Universidad Cat\'olica del Ecuador, Av. 12 de Octubre 1076 y Roca, Quito, Ecuador}

\begin{abstract}

%Recent decline in frog populations demand for large scale temporal and spatial studies to understand the phenomenon.
%by recording their calls in the wild. %
%large amounts of acoustic data already exist in archives which is  constantly increasing due to the appearance of automated programmable recorders. Manual sound analysis is no longer a practical option and 

%Machine Learning tools are being developed to assist researchers in this monumental task.
One non-invasive way to study frog communities is by analyzing long-term samples of acoustic material containing calls. This immense task has been optimized by the development of Machine Learning tools to extract ecological information.  We explored a likelihood-ratio audio detector based on Gaussian mixture model classification of 10 frog species, and applied it to estimate presence-absence in audio recordings from an actual amphibian monitoring performed at Yasun\'i National Park in the Ecuadorian Amazonia. A modified filter-bank was used to extract 20 cepstral features that model the spectral content of frog calls. Experiments were carried out to investigate the hyperparameters and the minimum frog-call time needed to train an accurate GMM classifier. With 64 Gaussians and 12 seconds of training time, the classifier achieved an average weighted error rate of 0.9\% on the 10-fold cross-validation for nine species classification, as compared to 3\% with MFCC and 1.8\% with PLP features. For testing, 10 GMMs were trained using all the available training-validation dataset to study 23.5 hours in 141, 10-minute long samples of unidentified real-world audio recorded at two frog communities in 2001 with analog equipment. To evaluate automatic presence-absence estimation, we characterized the audio samples with 10 binary variables each corresponding to a frog species, and manually labeled a sub-set of 18 samples using headphones.
%Therefore, it can be evaluated as binary classification performance in a sub-set of 18 human labeled samples. 
The one-vs-all Receiver Operating Characteristics curves were used to tune the likelihood-ratio detector per class in order to set operating points that minimize false positives while still allowing moderately noisy calls to be detected. A recall of 87.5\%  and precision of 100\% with average accuracy of 96.66\%  suggests good generalization ability of the algorithm, and provides evidence of the validity of this approach to study real-world audio recorded in a tropical acoustic environment.
%We found that it is not necessary to detect all the frog calls in the audio sample to achieve the goal of 10-species sub-set presence-absence estimation. As long as one call is detected is enough for presence, which is highly probable in ponds where frogs call repeatedly to attract mates. Nonetheless, for frogs that called once during the sample, a limitation of this approach was observed which was also the case during human annotation.  
Finally, we applied the algorithm to the available corpus, and show its potentiality to gain insights into the temporal reproductive behavior of frogs. %with figures of the quantity of detections and presence-absence over the samples of five months.

%For performance evaluation, we used a 10-fold cross validation routine to obtain the expected weighted error rate as quality measure. GMM with M = 2, 4, 8, 16, 32, 64 components were compared and the minimum amount of training segments was identified which yields a low error rate. Also, the algorithm was compared to the standard MFCC used in speech recognition observing an important improvement of about 2\% in its error rate. In addition, the approach was applied to study recordings performed in YNP yielding good correlation between the sound detections and the capture-recapture procedure performed after the recording.
\end{abstract}


\end{frontmatter}

%\linenumbers

\section{Introduction}\label{sec:introduction}

In long term ecological studies, it is important to quantify changes that occur on biodiversity and the ecosystem as a whole. Large scale temporal and spatial studies to understand the natural and anthropogenic induced population dynamics are demanded by the scientific community. In addition, recent anuran population declines around the world have motivated studies to gain an understanding of the phenomenon \cite{diechman}. One common way to obtain information is to assess frog communities since they are considered accurate indicators of environmental stress due to their aquatic and terrestrial habitat\cite{Simon2011,Relyea2005}. Researchers have been recording anuran audio signals in a labor intensive task that generates an ever increasing amount of audio data by using hand-held microphones, and networks of automated programmable recording equipment that stays in the field for months at a time\cite{Farina2017}. Manual analysis of hundreds of hours of audio is rather impractical and involves a long and tedious process~\cite{acevedo_using_2006} which has been aided in recent years by the use of modern software that relies on Machine Learning (ML) algorithms\cite{diechman}. Many important efforts have been made lately to provide a one-fits-all solution; however, evidence suggests that site and taxa specific algorithms are required to obtain the high levels of accuracy and reliability in automatic animal recognition systems necessary to extract ecological information \cite{XIE2016627,Towsey2012,Potamitis2014c,Ulloa2016,Bardeli2010}. Many frog call recognition approaches have been proposed in the literature, yet it remains unclear their suitability for the analysis of long audio samples recorded on the wild in places with high biological diversity using legacy recording equipment without a systematic approach. Years of frog-call data collection with variable quality audio samples remain unidentified and archived in audio repositories. Thus, the need for ML aid in site-specific frog species presence-absence estimation based on frog call detection in long audio recordings.

Male frogs use acoustic signaling for advertisement purposes to attract potential mates, defend their territory and show distress\cite{duellman_biology_1994}. Anuran vocalizations are commonly composed of a call that is formed by one or many sequenced notes also known as syllables. A syllable is an acoustic signal produced by air blown though vocal cords and resonated by a vocal sac~\cite[Chapter~4]{duellman_biology_1994}. A single call is chosen as the basic element for frog species detection since it exhibits heterospecific nature.  

Several studies found in the literature are focused on frog species automatic recognition. For instance, Brandes~\cite{brandes_feature_2008} introduced feature vectors extracted from spectrograms, and modeled calls of $10$ frogs recorded in the Amazon basin with hidden Markov models (HMM). Huan et al. ~\cite{huang_frog_2009} developed a frog sound identification system extracting $3$ features representing frog call syllables previously segmented reporting up to $90.3\%$ recognition rate using support vector machine (SVM) classification. The dataset consisted of $5$ species, 2 of which were clearly misclassified requiring further analysis. Lee et al.\cite{lee_automatic_2006} proposed a method using averaged Mel-frequency cepstral coefficients (MFCC) and linear discriminant analysis (LDA) to automatically identify $30$ types of frogs. Chen et al.~\cite{chen_automatic_2012} suggested a method based on pre-classification of syllable lengths, and a multi-stage averaged spectrum (MSAS) with template matching. This approach reported the best recognition rate on a dataset of 18 frog calls when compared to other methods based on dynamic time warping (DTW), k-nearest-neighbor (kNN) and SVM. Bedoya et al.~\cite{bedoya_automatic_2014} suggested an unsupervised methodology for automatic identification  based on a fuzzy classifier and MFCCs. The method was tested successfully with $13$ species of anurans found in Colombia. Aboudan et al.~\cite{aboudan_acoustic_2013} tested the ability of MFCC and linear predictive cepstral coefficients (LPCC) in the frog recognition process using Gaussian mixture models (GMM), but no real-world recordings containing frog calls was studied. Recently, an end-to-end Deep Neural Networks approach using convolutional neural nets (CNN) to classify spectrograms have been tried exhibiting 77\% classification accuracy showing a limitation in using that approach when a little amount of training data is available, which is normally the case with new species in the field. Xie et al. \cite{XIE2016627,XIE201713} proposed an intelligent system for estimating frog species richness and abundance that presented important results in long recordings made in Australia using a combination of acoustic features and random forests. These studies are a very important contribution to the state-of-the-art. However, none reported the application of their algorithms for the analysis of real-world, noise-contaminated audio recordings from an environment such as the Amazonian rainforest of Yasun\'i National Park (YNP) in eastern Ecuador. 

MFCCs have been applied in an out-of-the-box fashion, which exhibited limitations in their ability to model animal sound as reported in \cite{Towsey2012,cheng_call-independent_2010,fox_call-independent_2008}. This behavior is expected since the Mel-frequency filter bank used to generate MFCCs was designed based on the auditory properties of human hearing \cite{slaney_auditory_1998} and aims to model human voice. 

In this paper, we propose a modification to the Mel-scale filter-bank based on the spectral content of frog calls to obtain a modified cepstral feature set (m-FCC), and compare it experimentally to the performance of standard MFCC and PLP features sets used in speaker recognition. We performed experiments to find the minimum time of frog calls required to train accurate GMMs and investigated the hyperparameters of the models that minimize the error rate in the training-development set. In addition, a one-vs-all Receiver Operating Characteristics (ROC) analysis per class was performed to identify a threshold vector to allow a likelihood-ratio detector to reject sound segments that do not belong to the model set. The threshold is applied to control the sensitivity and specificity of the detection desired per class. For testing, we trained 10 GMMs of frog species using the labeled training-development dataset\footnote{An on-line demonstration is available at http://puceing.edu.ec:9001/Reconocimiento.aspx}, and applied those models to estimate frog species presence-absence in 141 (23.5 hours), 10-minute-long audio samples from a different distribution, with reduced quality, that was not used for training and validation of the algorithm. %The experimental results showed that the proposed feature set modification, with the experimentally identified GMM hyperparameters, performs well and yields high classification accuracy in the validation dataset. Moreover,
Performance evaluation in the practical presence-absence task validates the proposed approach in real-world conditions and prove the utility of the algorithm when unidentified acoustic data require analysis.    

%The main contributions are: {open data-set labeled, baseline for development, frog-call identification on-line website, a tool for long-recording analysis}

%In this work, we test the ability of likelihood-ratio frog call detection modeling probability density distributions by simple, but powerful GMMs. We compared widely used MFCC feature set with PLP features and propose a modified filter bank.  with GMM to recognize calls of $10$ frog species inhabiting the Yasun\'i National Park (YNP) in Ecuador. The recordings were made in the rainforest which is characterized by a complex acoustic environment. Thus, it is possible to find different sounds: birds, bats, crickets, mammals and other frog species sharing the spectrum at the same time. Experimental results demonstrated the effectiveness of the proposed method to achieve call-independent recognition on real recordings made in the wilderness. 

This paper is organized as follows. Section~\ref{sec:materials} describes the study site, the recording protocol used to register the frog calls in the wild, and the acoustic characteristics of the species available. Section~\ref{sec:methods} details the procedure followed for selection and annotation of the ground-truth dataset used in the experiments. Front-end segmentation is described and the modified cepstral features filter-bank presented. %Also, maximum likelihood classification and the log-likelihood-ratio detector applied are described, followed by the one-vs-all ROC analysis for the determination of the operating point desired per class. 
Section \ref{section:Experimental} is divided in two parts. The first one explains the experimental design and results of the parameter investigation. The second part describes the testing phase in real audio samples made by researchers in the wild. Finally, a discussion is presented in section~\ref{section:Discussion}, and conclusions summarized in section~\ref{section:Conclusion}.

\section{Materials}
\label{sec:materials}
\subsection{Study Site}
\label{subsection:stusite}
\begin{figure}[!ht]
	\centering
	\includegraphics[width = 1\textwidth]{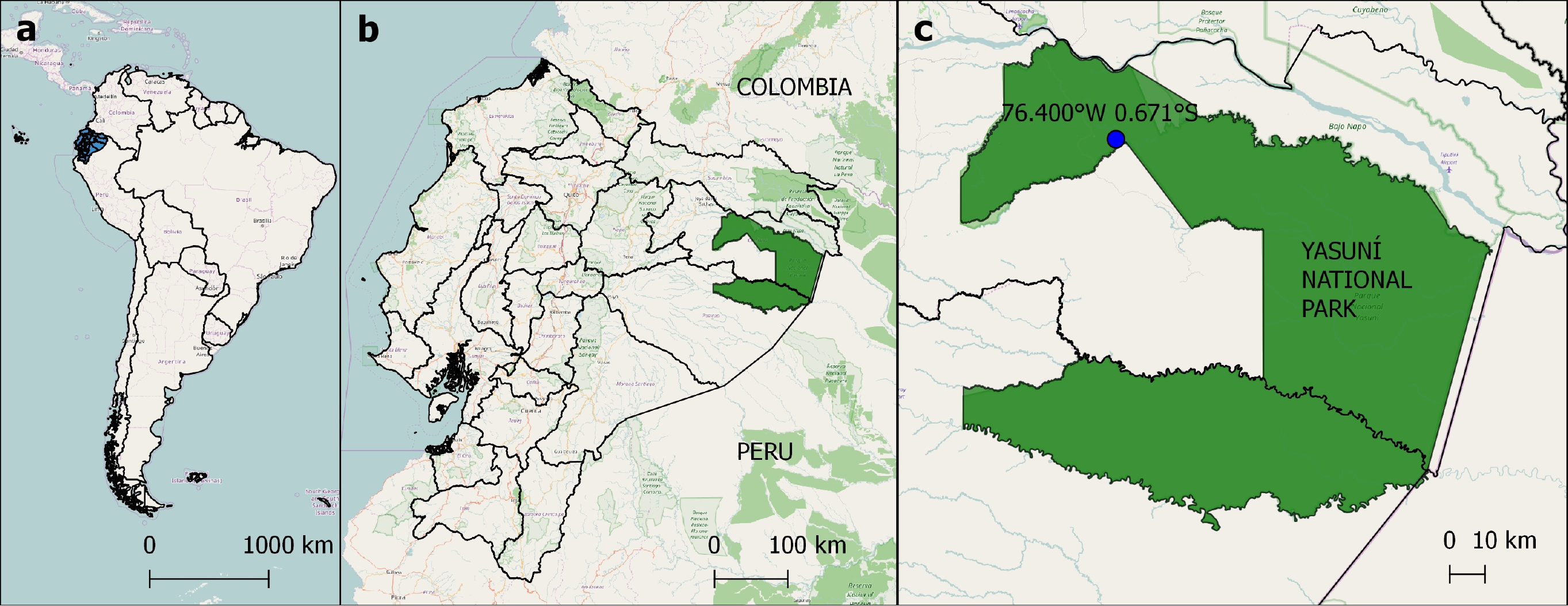}
	\caption{Location of the study area. a. Ecuador in South America. b. YNP in Ecuador. c.~PUCE's Yasun\'i Research Station in YNP.}
	\label{Fig:map}
\end{figure}
Frog calls were recorded within Yasun\'i National Park, which is located in the central eastern sector of the Ecuadorian Amazon region (\ang{1;05;0}{S}, \ang{75;55;0}{W}), in the provinces of \hbox{Orellana} and Pastaza (Figure \ref{Fig:map}). This \SI{9820}{\square\km} national park area is primarily rainforest that lies within the Napo moist forests ecoregion, and is considered one of the most biodiverse places on earth\cite{bass_global_2010}. It was designated as UNESCO Biosphere Reserve in 1989. Its climate is characterized by warm temperatures averaging \SI{24}{\degreeCelsius} to \SI{27}{\degreeCelsius} for all months; rainfall is high, approximately \SI{3200}{\mm} throughout the year. Relative humidity of YNP is between 80\% and 94\%. Average elevation of the park is low, from approximately \SI{190}{\m} to \SI{400}{m} above sea level; the territory is frequently crossed by hills of \SI{25}{\m} to \SI{70}{\m} high. Soil is mostly geologically young, product of fluvial sediments by the erosion of the Andes mountains~\cite{ministerio_del_ambiente_plan_2011}.

\subsection{Acoustic Environment}
\label{subsection:Aenvir}

% The Amazon basin has a tropical rainforest soundscape that is characterized by its high levels of sound diversity~\cite{riede_monitoring_1993}. This complex acoustic  environment is known to present a challenge for signal processing algorithms when automatic analysis of recordings is applied.~\cite{sueur2012global}. Since we focused on frog calls, we considered any other sound source in the field as noise.

\begin{figure}[!ht]
	\includegraphics[width=\textwidth]{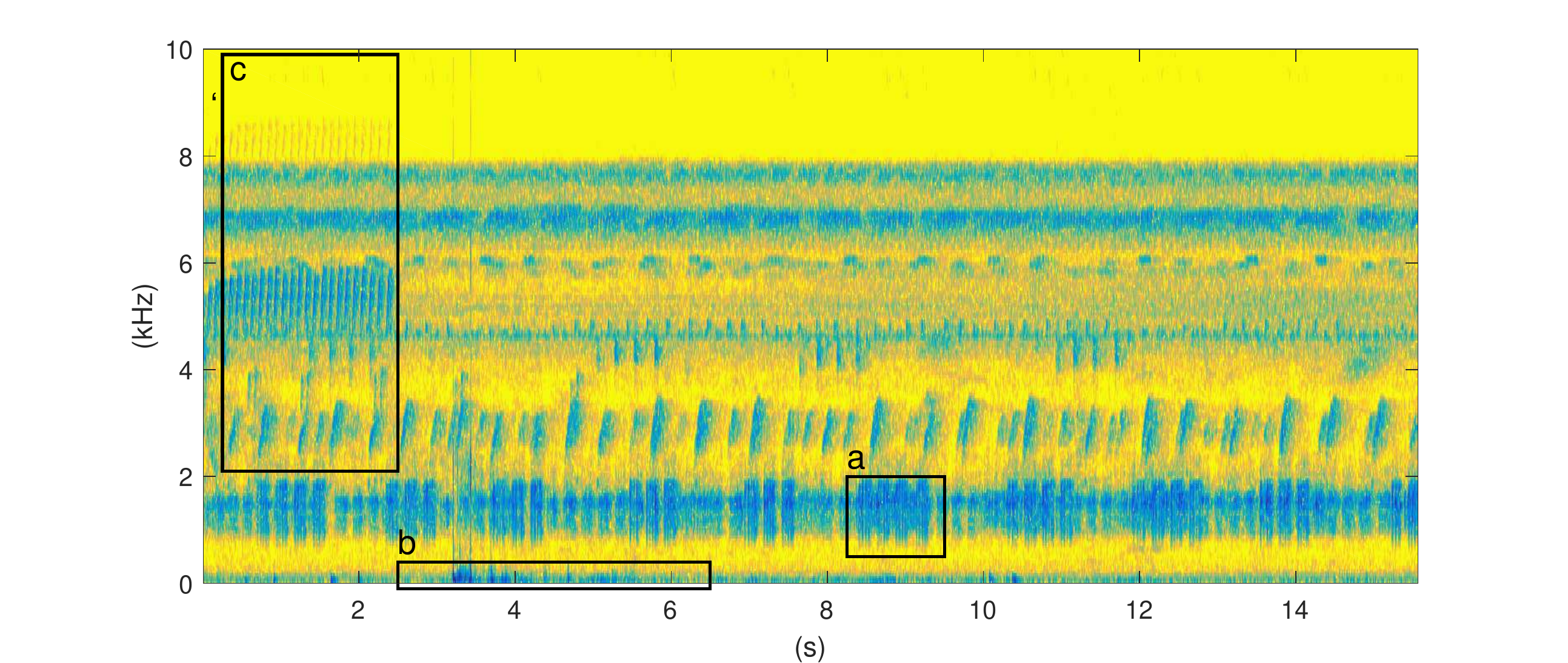}
	\caption{Spectrogram of a typical YNP soundscape showing a) frog call in a \textit{Rinhella margaritifera} chorus amidst b)antrophogenic and c)biotic noises.}
	\label{Fig:spectrogram}
\end{figure}

The acoustic environment of the Amazon basin is known to present a challenge for signal processing algorithms in automatic analysis of recordings~\cite{sueur2012global}. This region has a tropical rainforest soundscape with high sound diversity~\cite{riede_monitoring_1993}. In this paper, we focused on frog calls and any other sound source is considered noise. Previous studies~\cite{sueur2012global} have identified three main types of noise when recording soundscapes; namely, biotic noise, antrophogenic noise and environmental noise. A combination of these types of noise is present in the dataset used for this study. Some recordings contain antrophogenic noise like human voice and 60 Hertz "humming" of a nearby electric generator while others biotic noise from insects, mammals and nearby species. In addition, we identified noise of broad-band transient nature that resulted from friction of the microphone boom with the surrounding vegetation and water drops falling on the microphone while recording. Figure~\ref{Fig:spectrogram} shows the typical pond soundscape found within YNP in which a chorus of \textit{Rinhella margaritifera} amidst anthropogenic and biotic noise could be observed. 

\subsection{Acoustic Recording Protocol}
\label{section:DB}

The audio database containing frog calls used in this study was provided by Museo de Zoolog\'ia (QCAZ) of the Pontificia Universidad Cat\'olica del Ecuador (PUCE)~\cite{ron_amphibiawebecuador._2016}. The material was unlabeled and a few files contained only voice annotation made in the field.  For training and validation experiments, we used recordings made with a Sennheiser K6-ME67TM unidirectional microphone attached to digital recorders Olympus LS-10TM or Marantz PMD660TM with sampling frequencies of \SI{44100}{\Hz} and \SI{48000}{\Hz} at 16-bit resolution. Sound was archived in lossless WAV files in order to preserve the integrity of the audio. The recording schedule was from 7:00 p.m. to 2:00 a.m. at natural ponds and trails located within the YNP by different researchers during distinct sessions ranging from 2003 to 2013. Since locating the exact position of frogs calling in the wild at night is a difficult task, frog calls were registered aiming the microphone to the zone where the frogs were heard calling. Distance to the frog is therefore not available and varies within the dataset from a few meters to tens of meters for loud species. Considering that the distance to the frog is uncertain, we focused on the SNR when evaluating the detectability of a frog call in the sound file
\cite{estrella_selection_2017}. 

The audio used for testing was recorded at two ponds located close to PUCE's Yasun\'i research station by placing an omni-directional microphone 1.5 meters above the surface attached to a cassette recorder on a daily schedule during February (8 days), April (17 days), July (12 days), August (16 days), and September (13 days) of 2001. Pond 1 was recorded from 8:50 p.m. to 9:00 p.m, and Pond 2 from 1:50 a.m. to 2:00 a.m. The recordings were performed prior to a Visual Encounter Survey\cite{Padilla:Thesis:2005}.  The analog audio was transfered to digital audio in WAV format at 48000 kHz sampling rate using a USB digital audio converter in 2012. 

\subsection{Study Species}
\begin{figure}[!h]
\hspace*{-2cm}
	\includegraphics[width= 15 cm]{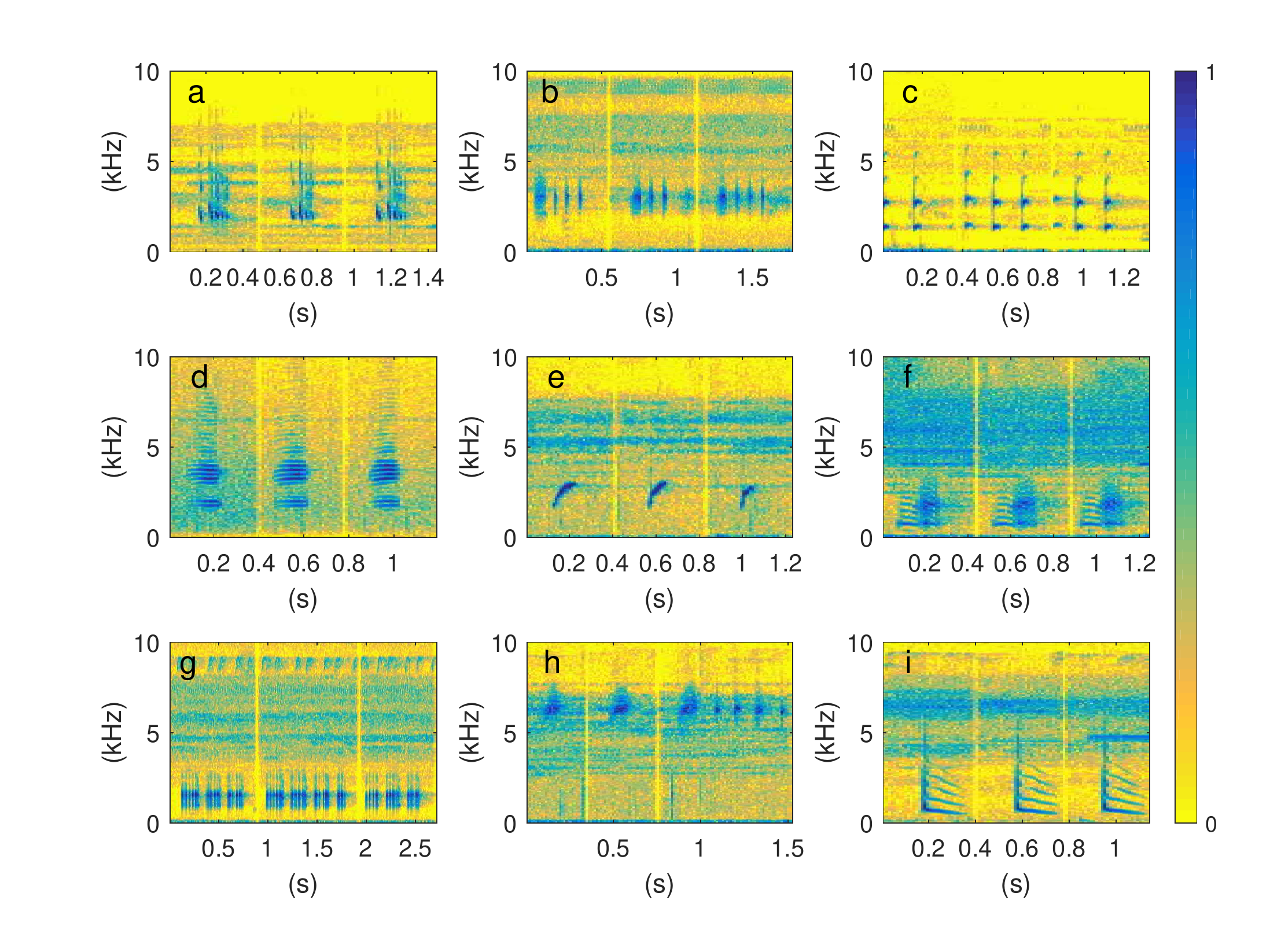}
	\caption{Spectrograms of 3 calls per species in the training-development set. a.\textit{Boana alfaroi} b.\textit{Dendropsophus bifurcus} c.\textit{Boana cinerascens} d.\textit{Pristimantis conspicillatus} e.\textit{Leptodactylus discodactylus} f.\textit{Boana lanciformis} g.\textit{Rhinella margaritifera} h.\textit{Dendropsophus parviceps} i.\textit{Engystomops petersi}}
	\label{Fig:spectrogram_frogs}
\end{figure}

We selected for our experiments the $10$ frog species listed in Table~\ref{Table:list}, which were chosen based upon availability at the time of labeling. Although more than 130 frog species have been identified so far in the study zone, typically only a few species are active at the same time and place. This is an important constraint for an automated analysis method since only a small subset of acoustic models are required for classification given a geographic location and timespan. To account for calls or sounds that are not modeled by the system, the option of unknown sound is included and its output can be studied by an specialist if necessary. Acoustic power in the frog calls of Table~\ref{Table:list} is mostly distributed in the range 430 to 7500 Hz depending on the species. Spectrograms of calls for each species with 1024 samples, 50\% overlap and Blackman-Harris window are shown in Figure~\ref{Fig:spectrogram_frogs}.  

\begin{table}[!h]
	\centering
	\caption{Study Species}
	
	\begin{tabular}{|c|c|c|c|c|c|}\hline
		Code&Species&\# of Calls&Seconds&Freq. range in Hz\\
		\hline
		$f_{01}$&\textit{Boana alfaroi}&98&44.7&1660 - 3100\\
        $f_{02}$&\textit{Dendropsophus bifurcus}&103&53.1&2300 - 3390\\
        $f_{03}$&\textit{Boana cinerascens}&169&105.5&1300 - 1530\\
        $f_{04}$&\textit{Pristimantis conspicillatus}&186&78&1630 - 3900\\
        $f_{05}$&\textit{Leptodactylus discodactylus}&330&146.3&1680 - 3260\\
        $f_{06}$&\textit{Osteocephalus fuscifacies}&50&23.52&1000 - 2500\\
		$f_{07}$&\textit{Boana lanciformis}&97&57.7&500 - 2720\\
		$f_{08}$&\textit{Rhinella margaritifera}&108&124.2&800 - 1700\\
		$f_{09}$&\textit{Dendropsophus parviceps}&38&40.1&5660 - 7500\\
		$f_{10}$&\textit{Engystomops petersi}&316&137.2&430 - 3140\\

		\hline
        \multicolumn{2}{|c|}{Total}&1495&810.32&\\
        \hline
	\end{tabular}
	
	\label{Table:list}
\end{table}

%\begin{figure}[t]
%\hspace*{-2cm}
%\includegraphics[width= 15 cm]{spectrogram_frogs_vg.eps}
%\caption{Spectrogram showing frog calls}
%\label{Fig:spectrogram_frogs_g}
%\end{figure}

\section{Methods}
\label{sec:methods}

\subsection{Frog-call Dataset}
\label{subsection:ASelec}

We generated a \textit{ground-truth} corpus of frog calls for training and testing ML algorithms\cite{estrella_selection_2016,dataset}. From the unlabeled audio provided by QCAZ museum of Zoology, we manually selected audio files containing frog calls. Nine species had enough acoustic material, from 40 to 146 seconds of calls, that allowed the creation of training-validation subsets used in the first set of experiments. Since classes were unbalanced, the split was made selecting 6,12 and 18 seconds of calls for training and the rest used for validation. $f_{06}$ was also included to generate a model for long recording analysis during testing.  Most frog calls were chosen with SNR higher than 3 dB, but we also included calls with background noise and some interference to study the performance of the algorithm in the noisy conditions that occur in the study zone. Field recordings containing human voice, mechanical artifacts or inter-specific overlapping calls were used neither to create the training-validation dataset nor to train the final GMMs. Table~\ref{Table:list} presents the number of calls and seconds of audio per each species available in the labeled dataset.  

Labeling of frog calls was aided by a short time energy (STE) based automatic segmentation algorithm described in section~\ref{section:segmentation}. Automatic segmentation of frog calls into syllables have been previously attempted by Jaafar et al.~\cite{jaafar_automatic_2013} with interesting results. The front-end STE segmentation algorithm outputs the start and end points of a  segment containing frog calls within the selected portion of audio. Each segment was manually labeled according to the species it belonged to by placing cue points signaling the start and end of the section containing calls. Automatic segmentation was preferred since an early attempt to perform manual endpoint selection resulted in lack of consistency among different annotators. 

For testing the algorithm in real world conditions, a subset of 18 audio samples of the 141 unidentified files was manually labeled by AET using headphones and spectrogram visualization. A vector of 10 binary variables (representing the species) was assigned per sample according to [$f_{01}$,$f_{02}$,$f_{03}$,$f_{04}$,$f_{05}$,$f_{06}$,$f_{07}$,$f_{08}$,$f_{09}$,$f_{10}$], in which one is presence and zero absence. For instance, [0 1 0 0 0 0 0 0 0 0], represents presence of \textit{D. bifurcus} and absence of all the others.
\subsection{Frog Call Segmentation}
\label{section:segmentation}

\begin{figure}[ht]
	\centering
	\includegraphics[width= 5 cm]{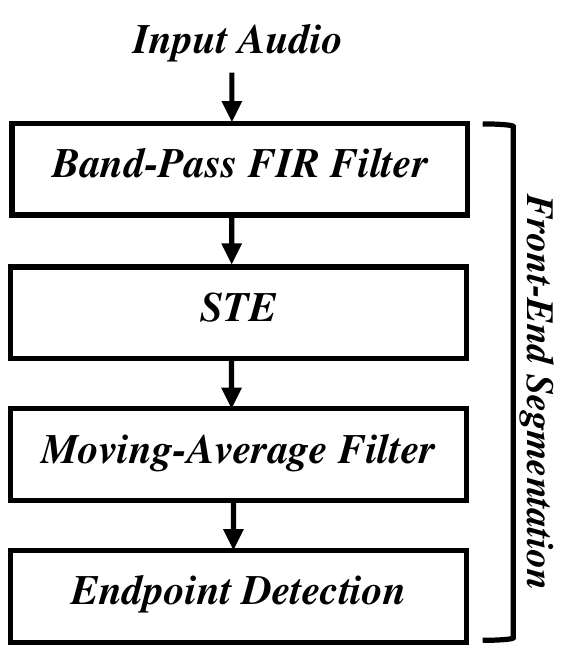}
	\caption{Frog call segmentation diagram.}
	\label{Fig:segmentation}
\end{figure}

Since a frog-call was chosen as the basic element of species identification, a segmentation technique that detects calls while avoiding portions of silence and noise was required for front-end processing. We adapted a classic voice analysis silence-removal method~\cite{rabiner_algorithm_1975} based on bandpass-filtering, STE estimation and thresholding. Figure ~\ref{Fig:segmentation} shows the algorithm pipeline.

First, the whole audio sample was divided into consecutive 30-second frames. A band-pass finite impulse response (FIR) filter was applied to the original audio signal. The cut-off frequencies are user defined and were chosen based on the frequency range spanning most of the call energy of the objective species. Table~\ref{Table:list} shows the frequency ranges of the filters used to generate the training set. The boundaries were calculated as the points were the spectral power is -20 dB relative to the point of maximum power of the frog call. It should be noted that the filter is applied only prior audio segmentation, but the original unfiltered audio is used for feature extraction and classification. Second, a STE sequence is generated from \SI{10}{\ms} consecutive frames with no-overlapping of the filtered signal~$s_{f}$ according to equation~\ref{eq:mSTE}.  
\begin{equation}
\label{eq:mSTE}
E(n)=\sum_{m=(n-1)N+1}^{nN} s_{f}(m)^{2},
\end{equation}
where $E(n)$ is the energy of frame $n$, $s_f(m)$ is the filtered discrete-time signal and $N$ is the number of samples of each \SI{10}{\ms} frame.

A moving-average (MA) filter was applied to the STE sequence to smooth transients and delimit STE of the whole frog call (or consecutive calls) instead of each separate note. The value of the MA = 12 was chosen empirically since it is related to the minimum frog-call duration that will be segmented.

%\begin{align}
%\label{eq:smooth}
%E_s(n)=\frac{1}{2Na+1}( E(n+Na)+E(n+Na-1)+\cdot\cdot\cdot \nonumber\\
%+E(n)+\cdot\cdot\cdot+E(n-Na)),
%\end{align}
%where $E_s(n)$ is a smoothed version of $E(n)$, $Na$ is the number of adjacent points in each size of E(n), and $2N+1$ is the total numbers of data points for the moving-average calculation. In our experiments a value of $Na=10$ proved sufficient to detect the frog calls in Table~\ref{Table:list}.

\subsection{Endpoint Detection}

The smoothed STE sequence was then transformed to \SI{}{\dB}, $STE_{dB}(n)$, and the following routine was applied to estimate the start and end points of the frog calls in the frame. 

\begin{enumerate}

\item Define a threshold value according to \begin{equation}\zeta_{dB}=\frac{max(STE_{dB}(n))-mean(STE_{dB}(n))}{C}\end{equation}. Where C is a constant determined empirically.

%	\item Define a threshold value according to \begin{equation}T=\frac{max(STE(n))-min(STE(n))}{C}\end{equation}. Where C is a constant determined empirically.
	\item If $3$ consecutive values of $STE_{dB}(n)$ are over the threshold, set a start-point. Subsequently, if $3$ consecutive values are below the threshold set the end-point.
\end{enumerate}

\begin{figure}[t]
	\includegraphics[width=\textwidth]{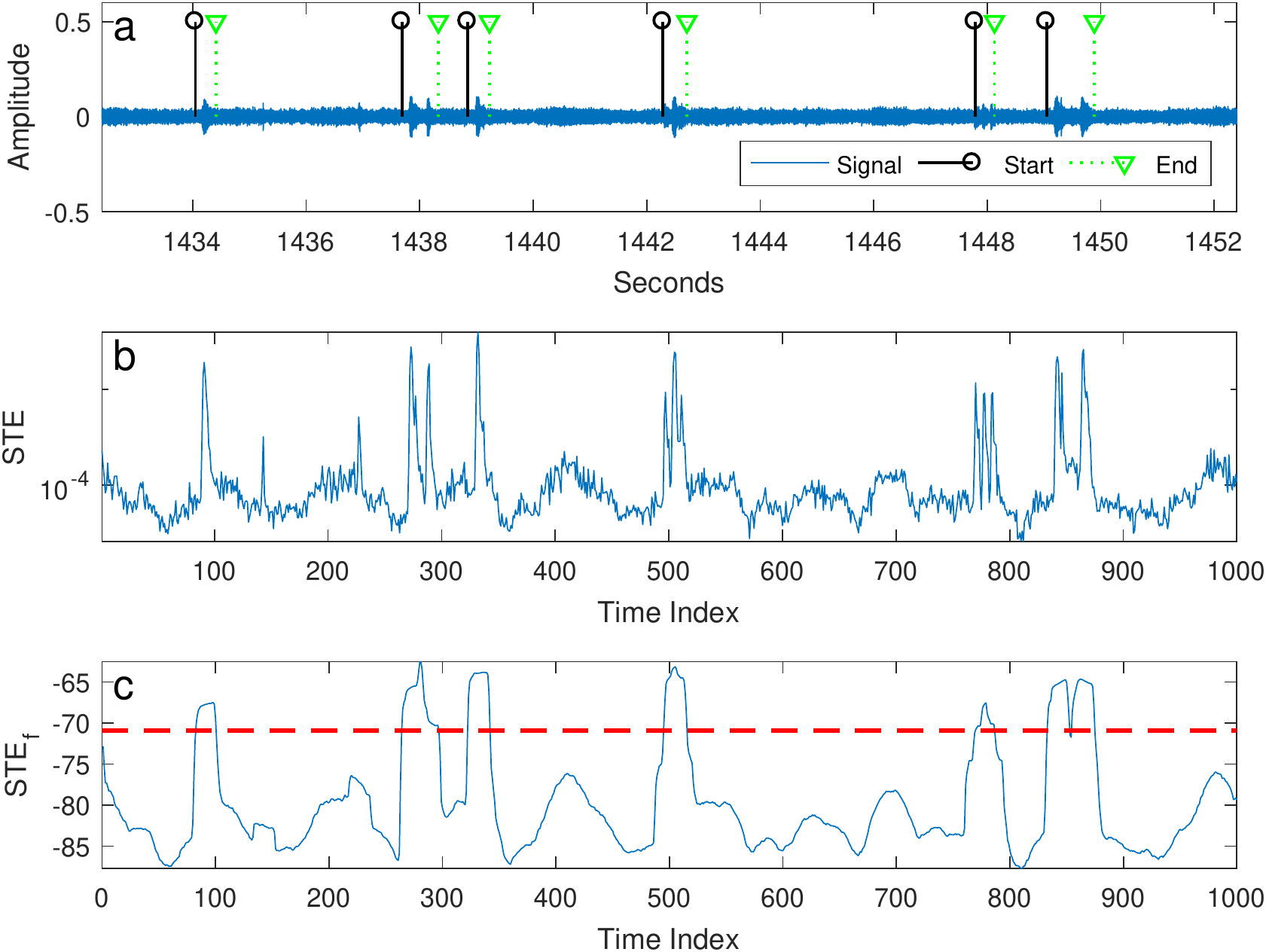}
	\caption{Segmentation of a \textit{Dendropsophus bifurcus} field recording. (a) 20 seconds of segmented input audio signal with start and end points. (b) STE sequence of 10 $ms$ frames of input signal (c) Smoothed STE in $dB$, $STE_f$ with threshold $\zeta_{dB}$ in dashed line.}
	\label{Fig:endpoint}
\end{figure}

The threshold allows for fine tuning the sensitivity of the endpoint detection. Its value is related to the SNR of the segmented audio that undergoes classification. Figure~\ref{Fig:endpoint} shows calls of \textit{Dendrophsophus bifurcus} detected applying the segmentation algorithm to a field recording.

\subsection{Acoustic features extraction}
\label{section:features}

\begin{figure}[!htb]
	\includegraphics[width=\textwidth]{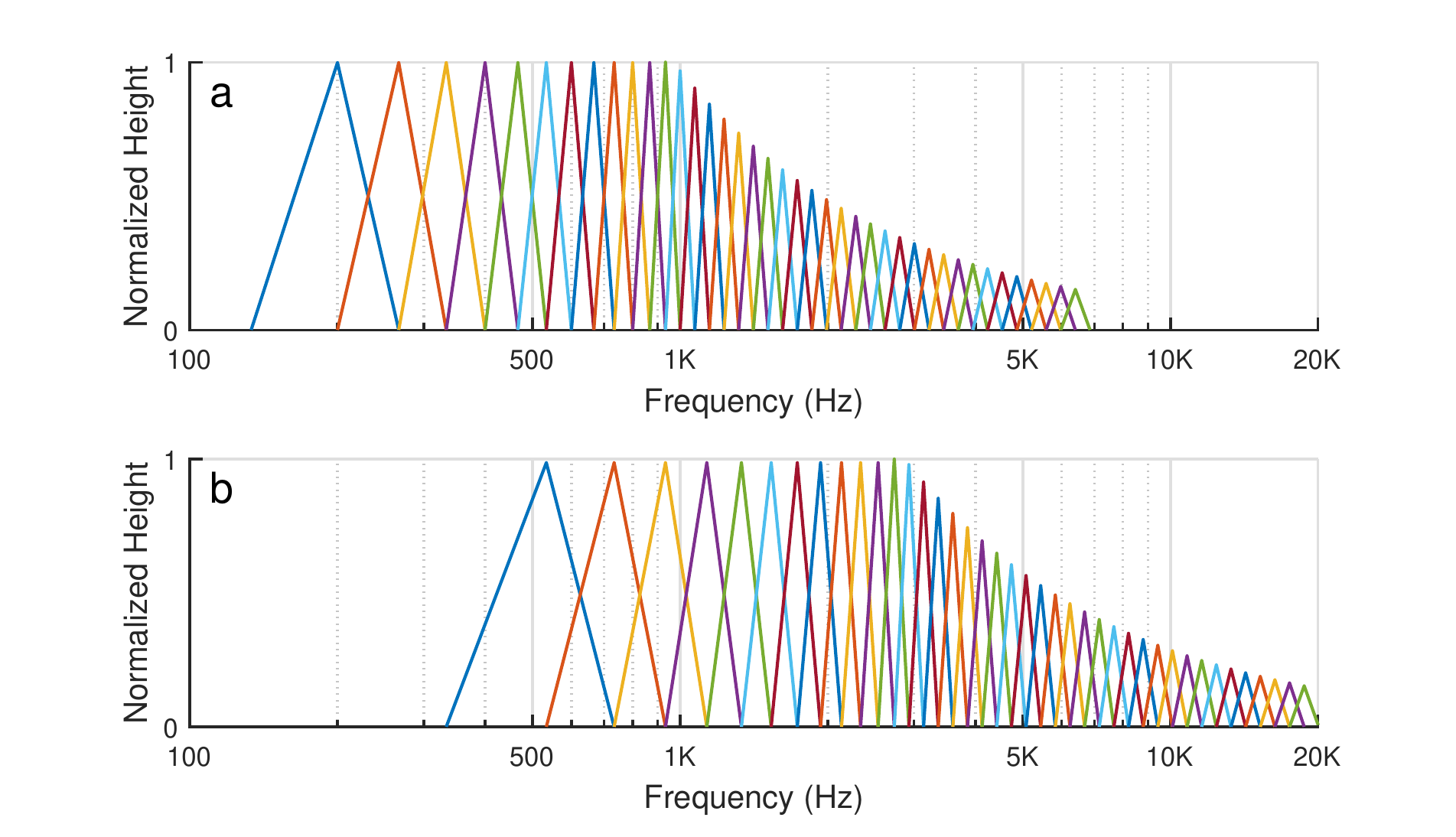}
	\caption{(a) layout of the mel-scale filter-bank (b)modified filter-bank proposed for frog-call identification}
	\label{Fig:filterbank}
\end{figure}

Mel-frequency cepstral coefficients (MFCC)~\cite{mermelstein1976distance} and perceptual linear predictive analysis (PLP)~\cite{Hermansky1992} have been the dominant feature sets used in automatic speech recognition systems~\cite{hermansky2013} as well as in automatic recognition of animal  sounds with interesting results in birds~\cite{fox_call-independent_2008,cheng_call-independent_2010}, odontocetes~\cite{roch_gaussian_2007}, anurans~\cite{bedoya_automatic_2014}. Those feature sets are optimized for human voice processing and have been applied mostly without modification to the problem of animal sound recognition obtaining important results. However, a close observation to the spectral energy of frog-calls reveal a different distribution than that of human voice. Therefore, it is not optimal to apply standard MFCC or PLP features without some modification to capture the spectral characteristics that differentiate frog calls. We propose using hand-crafted cepstral coefficients with a modified filter-bank distribution following the layout shown in Figure~\ref{Fig:filterbank} for frog-call recognition and compared its performance with standard MFCC and PLP-RASTA features. The procedure to extract the cepstral feature set is summarized as follows:

Fragments of sound containing frog-calls resulting from the segmentation step described in Section \ref{section:segmentation} were divided into 20 ms frames with 75\% overlap. Each frame was then pre-emphasized using the filter described by
\begin{equation}H(Z) = 1 - 0.99z^{-1}\end{equation} 
and Hamming-windowed to minimize discontinuities on the edges. The discrete Fourier transform (DFT) was taken and the triangular-shaped 40-element filterbank of Figure~\ref{Fig:filterbank}(b) was applied. 

The log of the energy of each filter was obtained and the discrete cosine transform (DCT) of the resultant vector of log-energies calculated to decorrelate the energies. Finally, the 20 first elements of the resultant vector were concatenated per each frame and the resultant matrix used as feature set for the classification step. A detailed description of Mel-cepstrum computation can be found in \cite{cheng_call-independent_2010,bedoya_automatic_2014,mermelstein1976distance}.

\subsection{Frog species model representation}

Gaussian mixture models have been shown to accurately model speaker identities when short-utterances of unconstrained speech are available for classification.~\cite{reynolds_robust_1995}. In the case of animal sound recognition, GMMs have been applied to identify vocalizations of individual birds\cite{cheng_call-independent_2010} and marine mammals~\cite{roch_gaussian_2007}. Since our aim is to identify the species of frogs calling in audio samples, we modeled each species by multi-variate GMMs. Formally,

A Gaussian mixture density is defined according to:

\begin{equation}
\label{eq:pxlamda}
p\left(\overrightarrow{x};\lambda\right)=\sum_{i=1}^{M}p_ib_i(\overrightarrow{x};\overrightarrow{\mu}_i,\Sigma_i),~~i = 1,2,...,M
\end{equation}

where $\overrightarrow{x}$  is a $D$-dimensional feature vector, $b_i(\overrightarrow{x};\overrightarrow{\mu}_i,\Sigma_i)$ are the component densities, and $p_i$ are the mixture weights that satisfy the constraint $\sum_{i=1}^{M}p_i=1$ \cite{reynolds_robust_1995}. Each component density is a Gaussian function of $D$ variables:

\begin{equation}
\label{eq:bix}
b_i(\overrightarrow{x};\overrightarrow{\mu}_i,\Sigma_i)=\frac{1}{(2\pi)^{D/2}|\Sigma_i|^{1/2}} \exp \left\{ -\frac{1}{2} (\overrightarrow{x}-\overrightarrow{\mu}_i)^{T} \Sigma_i^{-1} (\overrightarrow{x}-\overrightarrow{\mu}_i)\right\},
\end{equation}

with mean vector $\overrightarrow{\mu}_i$ and covariance matrix $\Sigma_i$. Each GMM  is denoted by its mean vector, covariance matrix and the mixture weights according to:

\begin{equation}
\label{eq:lamda}
\lambda=\left\{ p_i,\overrightarrow{\mu}_i,\Sigma_i \right\},~i=1,...,M.
\end{equation}

We need a model $\lambda$ for each species available in the labeled dataset of Table~\ref{Table:list}. Thus, we generated a set of training vectors for every utterance in the dataset by extracting cepstral features at fixed time steps. The resultant matrix $X = (\overrightarrow{x_1}, \overrightarrow{x_2},...,\overrightarrow{x_n})$ was then used to train models $\lambda_1$, $\lambda_2$,...,$\lambda_{10}$ for each species by maximum likelihood estimation (MLE) using the expectation-maximization (EM) algorithm described in~\cite{dempster1977maximum,reynolds_robust_1995,cheng_call-independent_2010}. The model was initialized by setting the mean values using the k-means++ algorithm~\cite{arthur2007k}, the initial covariance matrix was set as diagonal with element $(j,j)$ as the variance of $X(:,j)$, and the initial mixing proportions were set as uniform.  

%\begin{figure}[htb!]
%	\centering	
%	\includegraphics[height=5 cm]{training.eps}
%	\caption{Training process diagram.}
%	\label{Fig:training}
%\end{figure}

\subsection{Frog species identification}

%\begin{figure}[bth!]
	%\centering
	%\includegraphics[height=5cm]{identification.eps}
	%\caption{Identification process diagram.}
	%\label{Fig:identification}
%\end{figure}

The output of front-end segmentation yields a sequence of vectors $X=\{\overrightarrow{x}_1, ..., \overrightarrow{x}_n   \}$ that contain characteristics of the sound source that generated it. We need to find the species model with the maximum \textit{a posteriori} probability for $X$. Formally,

%The identification procedure of Figure~\ref{Fig:identification} was applied to each isolated call $w$, $1\leq w \leq N_C$. Since for this project we required to identify the $10$ species of YNP in Table~\ref{Table:list}, a set of ten frog species $F = \{f_{01}, f_{02},..., f_{10}\}$ was established. Each frog species is represented by a model $\lambda_k$, $k=1, 2, ..., 10$. The goal is to find the frog model which has the maximum posterior probability for an input sequence $\overrightarrow{X}=\{\overrightarrow{x}_1, ..., \overrightarrow{x}_T   \}$. For this study, the input sequence is a matrix of MFCC coefficients of each call $w$ of the audio signal. Minimum error Bayes's decision procedure was applied:

\begin{equation}
\label{eq:hatf}
\hat{f}=\arg \max_{1\leq k \leq S}{Pr}(\lambda_k|X)=\arg \max_{1\leq k \leq S} \frac{p(X|\lambda_k)}{p(X)} {Pr}(\lambda_k),
\end{equation}

where $\hat{f}$ is the hypothesized frog species and S is the number of models. Even though prior probabilities could be defined based on the geographic location of the study, we assume identical prior probabilities of frog species ${Pr}(\lambda_k)$ and remove $p(X)$ since it is the same for all models. The decision rule becomes:

\begin{equation}
\label{eq:hatf1}
\hat{f}=\arg \max_{1\leq k \leq S}  p(X|\lambda_k).
\end{equation}

Since front-end processing produces variable size audio segments, we need to normalize for size $T$. Applying logarithms and assuming independence between observations, the species identification stage calculates:

%\begin{equation} 
%\label{eq:hatf2}
%\hat{f}=\frac{1}{T}\arg \max_{1\leq k \leq S}  \sum_{t=1}^{T}  \log~p(\overrightarrow{x}_t|\lambda_k).
%\end{equation}

%Each audio segment on the input $w$ is composed of $T$ blocks. So, the expression $\sum_{t=1}^{T}  \log~p(\overrightarrow{x}_t|\lambda_k)$ in Eq.~\ref{eq:hatf2} is dependent on $T$ value. Normalizing Eq.~\ref{eq:hatf2} based on $T$ value, we have:

\begin{equation}
\label{eq:hatf3}
\hat{f}=\arg \max_{1\leq k \leq S} \frac{1}{T} \sum_{t=1}^{T}  \log~p(\overrightarrow{x}_t|\lambda_k).
\end{equation}

%The value of $\max \left[ \frac{1}{T} \sum_{t=1}^{T}  \log~p(\overrightarrow{x}_t|\lambda_k) \right]$, $1\leq k \leq 10$, is interpreted as the ML of the model $\lambda_k$ that best matches the input signal. However,

\subsection{Species detection}

With the hypothesized frog species $\hat{f}$, we need to determine if the sound segment was actually produced by $f$. To accomplish this goal, we applied the log-likelihood ratio statistic defined by: 

\begin{equation}
\label{eq:loglike}
\Lambda(X)=\log~p(X|\lambda_{hyp})-\log~p(X|\lambda_{\overline{hyp}})
\end{equation}

where $\Lambda(X)$ is a score that informs how likely it is for a given call segment to belong or not to the hypothesized species model $\lambda_{hyp}$ that represents $f$, and $p(X|\lambda_{\overline{hyp}})$ is the probability density function of the set of alternative species in the model set.  This task is known as verification in the speaker detection literature\cite{Reynolds2000}. Although more than one species usually call at the same time, for the application of presence-absence estimation we focused on single-species detection per segment. A close observation to the calling patterns of frogs during reproduction reveals that frogs call repeatedly in choruses sharing the time and frequency resources available. For a 10-minute sound sample, we assumed that at least one segment of sound is single-species composed.  To represent the pdf of the alternative species model we applied the \textit{median()} function of the set of non hypothesized models in the set $S$:

\begin{equation}
\label{eq:loglikealt}
p(X|\lambda_{\overline{hyp}})= median\{p(X|\lambda_{f_0}),p(X|\lambda_{f_1},...,p(X|\lambda_{f_{S-1}}))\}
\end{equation}

To find a threshold of the score $\Lambda(X)$ that permits detection of frog calls with high likelihood while rejecting non relevant sound segments (overlapped calls, noise, non modeled species, etc), we defined threshold values per class applying one-vs-all receiver operating characteristic (ROC) analysis.

%\begin{equation}
%H_0: X \text{ is from the hypothesized species } \lambda_k 
%\end{equation}
%and
%\begin{equation}
%H_1: X \text{ is \underline{not} from the hypothesized species } \lambda_k 
%\end{equation}

%The likelihood ratio test is given by 
%\begin{equation}
%\frac{p(X|H_0)}{p(X|H_1)} 
%\begin{cases}
%\geq\theta\text{~~~~accept~}H_0\\
%<\theta\text{~~~~reject~}H_0,\\ 
%\end{cases}   
%\end{equation}

%\begin{equation}
%\label{eq:maxlikelihood}
%\max_{1\leq k \leq 10} \frac{1}{T} \sum_{t=1}^{T}  \log~p(\overrightarrow{x}_t|\lambda_k) \geq \gamma_k.
%\end{equation}

\subsection{Threshold vector selection}
\begin{figure}[!htb]
\centering
	\includegraphics[width=1\textwidth]{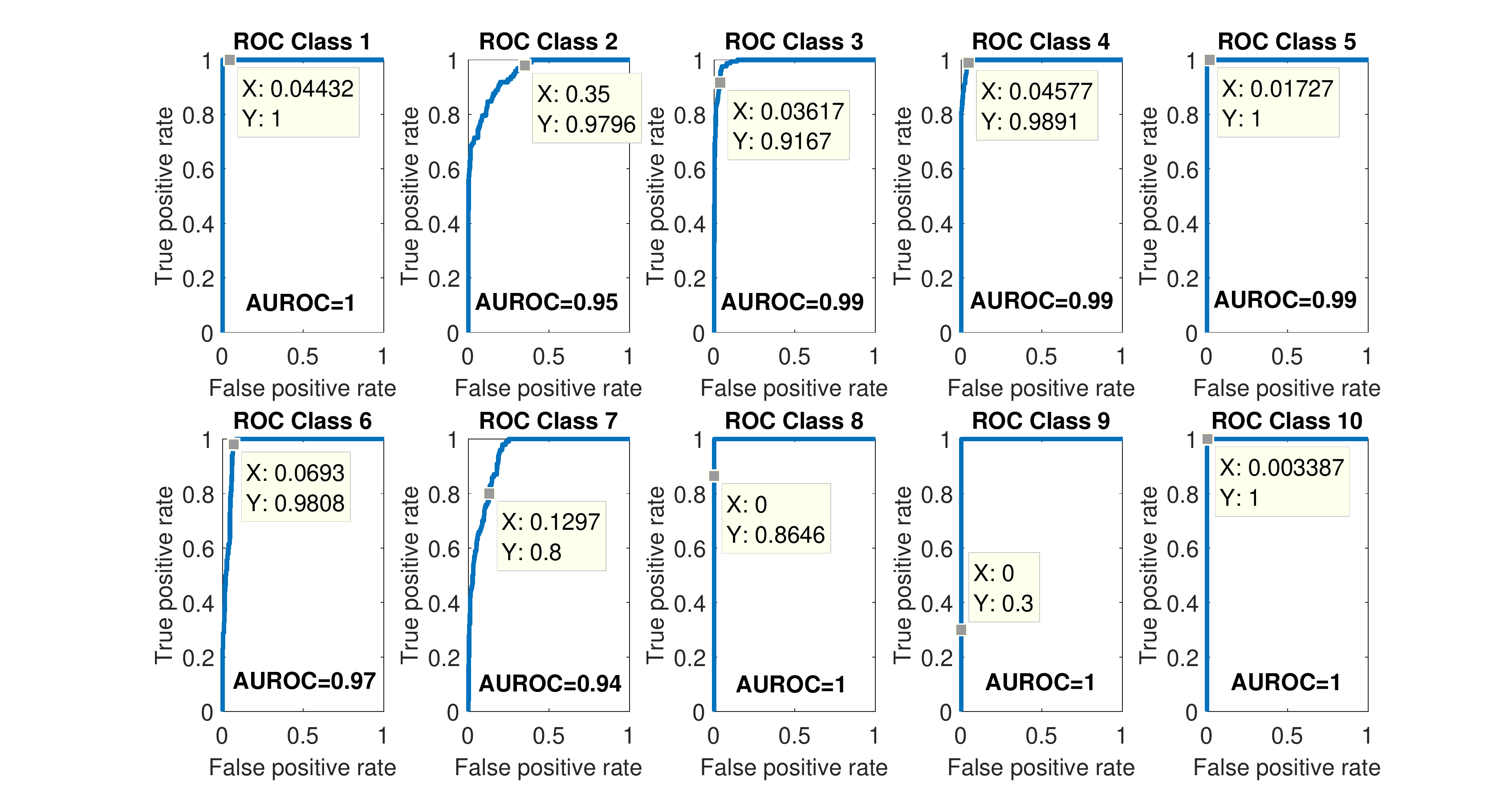}
	\caption{Per class Receiver operating characteristics curves and the chosen operating points.}
\label{Fig:Rocurves}
\end{figure}
ROC curves~\cite{hanley1982meaning,Fawcett2005} for the multi-class detection problem are shown in Figure \ref{Fig:Rocurves}. A one-vs-all approach was used to calculate the true positive rate (TPR) versus the false positive rate (FPR) for each class in relation to a varying log-likelihood ratio score. To consider an acoustic event as a class detection, its score value must be above a defined threshold per class. Higher threshold values enable a selective detection while lower values increase the sensitivity. A trade-off between false detections and false rejections exists and the operating point depends on the application. In the case of frog call presence-absence estimation in long recordings it is desirable to minimize false alarms while maintaining a TPR as high as possible. Since the audio recorded in YNP was registered with a unique microphone in a fixed site, it is important to detect all the frog calls possible in the range of the microphone. Therefore, the operating point requires high TPR while maintaining FPR reasonably low. Some intuition of the behavior and frequency of calling per species is desirable when setting the threshold. We began with a fixed threshold vector of operating points corresponding to FPR of 5\% for all the species and fine tuned the thresholds comparing the analysis output to the manually annotated corpus. The final threshold vector used in the analysis of section \ref{subsection:Field} was [5 3 6 9 6.75 5.25 5.5 11 6 6] and the corresponding operating points are shown in Figure \ref{Fig:Rocurves}.      

\section{Experimental Results}
\label{section:Experimental}

We divided this section in two parts. First, a set of experiments was designed to identify the hyperparameter values that perform best in recognizing the species of frogs from a frog call in the development dataset as well as the minimum time required to train accurate GMMs. Second, the parameters obtained in the first part were used to train production GMMs using all the material available in the dataset. Automatic analysis of real field recordings coming from a different distribution was performed, and its results compared to human level performance in the presence-absence task.   

\subsection{Parameter investigation}

\begin{figure}[!h]
	\centering
	\includegraphics[width=0.75\textwidth]{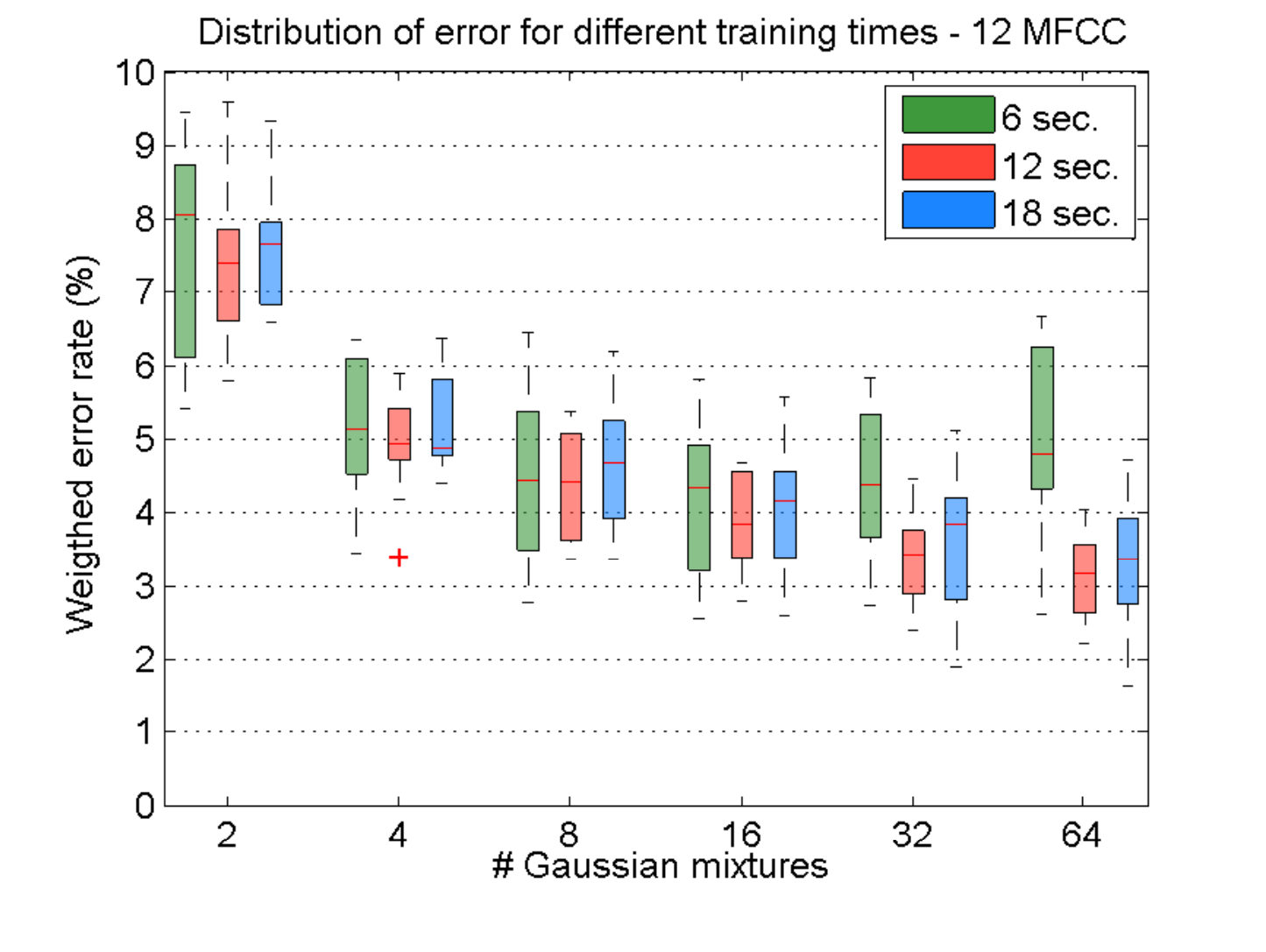}
	\caption{Weighted error rate for different training times.}
	\label{Fig:weighted_error_time}
\end{figure}

The number of components $M$ in a mixture needed to model frog species adequately, and the minimum training time required were determined by the following experiment. Nine frog models with 2, 4, 8, 16, 32 and 64 component Gaussian densities and diagonal covariance matrix were trained using 6, 12, 18 seconds of frog-call corresponding to 600, 1200 and 12800 12-dimensional mel-cepstral feature vectors. The dataset was divided into a training set of 6, 12 and 18 seconds and the remaining calls were used for testing. We applied 10-fold cross-validation with random segment selection on all the dataset per each species to model the distribution of the weighted error rate (WER). The WER was calculated as the average on the individual per species Bayesian error rate of the nine species to account for the unbalanced classes. Figure~\ref{Fig:weighted_error_time} shows the distribution of WER for different training times and number of mixtures M for the 10-fold cross-validation procedure.    

The following observations can be made from the results. For the 3 training times tested, there was an increase in identification performance from 2 to 16 mixture components leveling off for 32 and 64 components. However, when training with 6 seconds the identification performance degrades with increasing model order suggesting that at least 12 seconds of training set and more than 32 mixture components are required to model the frog species adequately. This result is important since it provides a guideline for the lower limit of training set required when training frog species GMMs. %Second, the WER distribution is narrower when 12 seconds were used as compared to 18 seconds of training time. This results suggests that more training time results in over-fitting.  

%\begin{equation}
%\label{eq:successfulrate}
%{Success~rate}(\lambda_k)=\frac{{Calls~successfully~ %recognized~}({Frog~}k)}{{Total~calls~}({Frog~}k)}.
%\end{equation}

\begin{table}[!hbt]
	\caption{Frog call recognition error rate per species with respect to the number of component densities for 20 m-FCC.}
    \label{Table:summary}
   \begin{tabular}[t]{|c|c|c|c|c|c|c|}\hline
	Frog&\multicolumn{6}{c|}{Model Order}\\\cline{2-7}
    Species&M = 2&M = 4&M = 8&M = 16&M = 32&M = 64\\
	\hline
	%&    &    &    &$M=1$&$100.00$\\*
	%&    &    &    &$M=2$&$100.00$\\*
	%&    &    &    &$M=4$&$100.00$\\*
	\textit{H. alfaroi}~$(\%)$&$0$&$0$&$0$&$0$&$0$&0\\*
	%&    &    &    &$M=16$&$100.00$\\*
	%&    &    &    &$M=32$&$100.00$\\*
	%&    &    &    &$M=64$&$100.00$\\*
	\hline
	%&    &    &     &$M=1$&$96.59$\\*
	%&    &    &     &$M=2$&$97.72$\\*
	%&    &    &     &$M=4$&$100.00$\\*
	\textit{D. bifurcus}~$(\%)$&$9.4$&$6.7$&$2.7$&$0.8$&$0.9$&0.8\\*
	%&    &    &     &$M=16$&$100.00$\\*
	%&    &    &     &$M=32$&$100.00$\\*
	%&    &    &     &$M=64$&$100.00$\\*
	\hline
	%&    &    &    &$M=1$&$89.18$\\*
	%&    &    &    &$M=2$&$89.18$\\*
	%&    &    &    &$M=4$&$89.18$\\*
	\textit{H. cinerascens}~$(\%)$&$0.4$&$0.2$&$0.2$&$0.1$&$0.1$&0.2\\*
	%&    &    &    &$M=16$&$89.18$\\*
	%&    &    &    &$M=32$&$89.18$\\*
	%&    &    &    &$M=64$&$89.18$\\*
	\hline
	%&   &    &    &$M=1$&$62.50$\\*
	%&   &    &    &$M=2$&$68.75$\\*
	%&   &    &    &$M=4$&$87.50$\\*
	\textit{P. conspicillatus}~$(\%)$&$0.5$&$0.6$&$0.3$&$0.3$&$0.3$&0.3\\*
	%&   &    &    &$M=16$&$87.50$\\*
	%&   &    &    &$M=32$&$93.75$\\*
	%&   &    &    &$M=64$&$100.00$\\*
	\hline
	%&    &    &    &$M=1$&$100.00$\\*
	%&    &    &    &$M=2$&$100.00$\\*
	%&    &    &    &$M=4$&$100.00$\\*
	\textit{L. discodactylus}~$(\%)$&$0.3$&$0.2$&$0.3$&$0.1$&$0.1$&0.2\\*
	%&    &    &    &$M=16$&$100.00$\\*
	%&    &    &    &$M=32$&$100.00$\\*
	%&    &    &    &$M=64$&$100.00$\\*
	\hline
	%&   &    &    &$M=1$&$100.00$\\*
	%&   &    &    &$M=2$&$100.00$\\*
	%&   &    &    &$M=4$&$100.00$\\*
	\textit{B. lanciformis}~$(\%)$&$7.8$&$3.4$&$1.8$&$1.6$&$1.3$&1.6\\*
	%&   &    &    &$M=16$&$100.00$\\*
	%&   &    &    &$M=32$&$100.00$\\*
	%&   &    &    &$M=64$&$100.00$\\*
	\hline
	%&    &     &     &$M=1$&$91.00$\\*
	%&    &     &     &$M=2$&$91.00$\\*
	%&    &     &     &$M=4$&$94.00$\\*
	\textit{R. margaritifer}~$(\%)$&$11.7$&$6.1$&$5.2$&$4.2$&$4.1$&3.2\\*
	%&    &     &     &$M=16$&$99.00$\\*
	%&    &     &     &$M=32$&$99.00$\\*
	%&    &     &     &$M=64$&$99.00$\\*
	\hline
	%&    &    &    &$M=1$&$75.86$\\*
	%&    &    &    &$M=2$&$82.76$\\*
	%&    &    &    &$M=4$&$93.10$\\*
	\textit{D. parviceps}~$(\%)$&$4.7$&$3$&$3.7$&$3$&$3.4$&3\\*
	%&    &    &    &$M=16$&$93.10$\\*
	%&    &    &    &$M=32$&$93.10$\\*
	%&    &    &    &$M=64$&$93.10$\\*
	\hline
	%&    &     &     &$M=1$&$50.00$\\*
	%&    &     &     &$M=2$&$71.91$\\*
	%&    &     &     &$M=4$&$86.30$\\*
	\textit{E. petersi}~$(\%)$&$0.2$&$0$&$0$&$0$&$0$&0\\*
	%&    &     &     &$M=16$&$93.15$\\*
	%&    &     &     &$M=32$&$94.52$\\*
	%&    &     &     &$M=64$&$95.89$\\*
	\hline
	%&    &    &    &$M=1$&$57.14$\\*
	%&    &    &    &$M=2$&$91.83$\\*
	%&    &    &    &$M=4$&$100.00$\\*
	%$f_{10}$&$23$&$49$&$72$&$M = 8$&$100.00$\\
	%&    &    &    &$M=16$&$97.95$\\*
	%&    &    &    &$M=32$&$100.00$\\*
	%&    &    &    &$M=64$&$97.95$
   	%\hline
\end{tabular}
\end{table}

Finally, in Table \ref{Table:summary} the error rate per species is presented according to the number of Gaussian mixture components used to model each species with 12 seconds of training time and 20-mFCC. Variable error rates across distinct M suggests that the optimal model order is not the highest for each species in the data set. A different model order could be chosen for each species to avoid over-fitting as suggested by Cheng in \cite{cheng_call-independent_2010}. However, we selected model order M = 64 that gives the minimum error rate on the average since it is not clear how to choose the optimal model order for species that are not included in the training - validation set, but could be included in the future.     

\subsection{Features comparison}

\begin{figure}[!htb]
	\centering
	\includegraphics[width=0.75\textwidth]{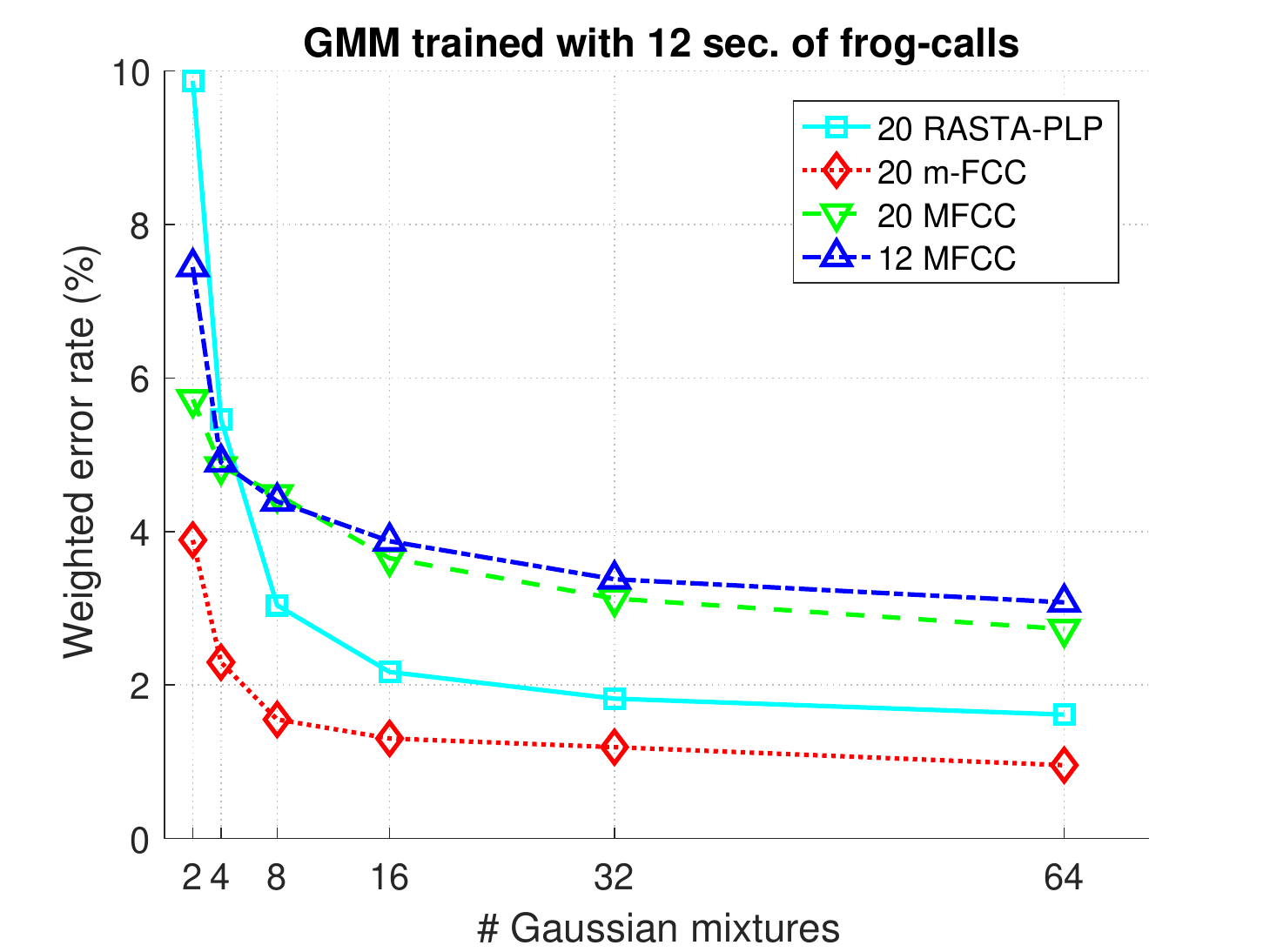}
	\caption{Weighted error rate for GMMs trained with different cepstral features.}
	\label{Fig:weighted_err_feat}
\end{figure}

In order to investigate the WER with respect to the feature set used to model the frog calls, GMMs were trained using 12 MFCC, 20 MFCC, 20 PLP-RASTA and 20 modified-filterbank cepstral features(m-FCC). The results are shown in Figure~\ref{Fig:weighted_err_feat} which presents the WER with respect to the number of mixture components for each feature set. Standard MFCC used in automatic speaker recognition exhibited the lowest performance with a slight improvement when the number of features increased from 12 to 20. Additionally, 19th PLP-RASTA outperformed standard MFCC performance by approximately 1\% when more than 16 mixture components were used suggesting that the Bark filter-bank used to calculate the PLP feature set allows the spectral characteristics of the frog calls to be captured better than MFCC. Finally, classification performance of the 20 cepstral features calculated using the modified filter-bank described in Section 3.4 surpass the others. The results suggests that the modification to the filter-bank in order to model the spectral shapes frog calls rather than those of human voice is appropriate. 

\subsection{Field recordings analysis}
\label{subsection:Field}

\begin{figure}[!h]
	\centering
	\makebox[\textwidth][c]{\includegraphics[width=1.66\textwidth]{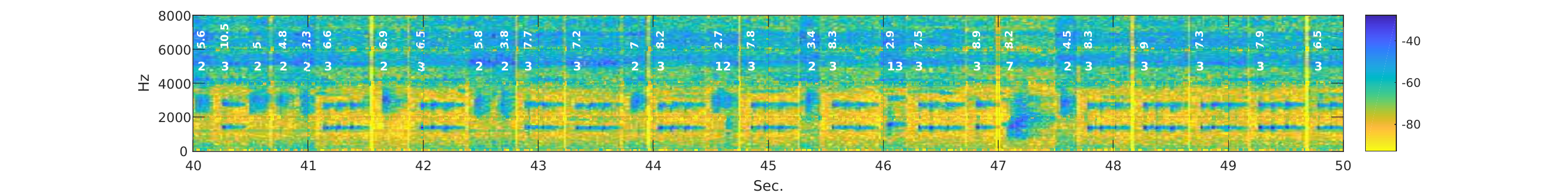}}
	\caption{Snippet of frog call detections. The numbers in the first line are the likelihood-ratio scores for each segment. Below is the species code \textit{D.bifurcus} (2), \textit{B.cinerascens} (3), \textit{B.lanciformis} (7). (12) and (13) are classifications that did not pass the set threshold.}
	\label{Fig:snippetclass}
\end{figure}

Ten GMMs were trained using all the labeled dataset, and applied to analyze 23.5 hours of audio in 50 WAV files with focus on presence-absence estimation. Each file contained three 10-minute samples delimited using cue points to inform its position to the algorithm. The audio contains unidentified calls amidst different types of noise and distortion resultant from volume variation during recording, clipping, malfunctioning cable, microphone friction, rain and digitalization noise. Scanning each file took 50 seconds approximately with a laptop running a 2.6 GHz processor and 16 GB RAM. Figure \ref{fig:longrecordexample} shows a 40-second snippet of segmentation stage, and classified segments with their respective likelihood-ratio score are presented in Figure \ref{Fig:snippetclass}.

Presence-absence estimation results of all the corpus are summarized in Figure 11 for both ponds studied. One binary vector per sample was obtained and plotted per day and month according to the sampling schedule available in the 5-month period. To evaluate its performance, the 18 audio samples of February were manually annotated by AET using headphones and spectrogram visualization. The resultant 10-variable binary vectors were compared to the output of the algorithm variable by variable, and binary classification performance metrics were applied. The resultant scores are presented in Table \ref{Table:perform} %We calculated recall and precision, specificity, F1-score, Matthews correlation coefficient (MCC) and accuracy. T

\begin{table}[!htb]
\centering
	\caption{Performance measures for the presence-absence task}
    \label{Table:perform}
   \begin{tabular}[t]{|c|c|}\hline
Metric&Score\\
\hline
Recall   &    0.875\\
Precision &   1\\
F1       &    0.933\\ 
MCC     &     0.914\\
Specificity & 1\\
Accuracy   &  0.966\\
\hline
\end{tabular}
\end{table}

Calls of \textit{B. alfaroi}, \textit{P. conspicillatus}, \textit{R. margaritifer} and \textit{D. parviceps} did not exist in the recordings and were correctly estimated as absent by the machine with the exception of one sample in July in which a false detection of \textit{P. conspicillatus} occurred. In contrast, \textit{D. bifurcus} was detected in all the annotated samples by both human and machine. The algorithm was able to estimate absence correctly for the species that did not call during the sample while presence of \textit{O. fuscifacies} and \textit{B. lanciformis} presented a challenge. Those species called only once or twice per sample, making it difficult to detect them during manual screening as well as during automatic analysis. Nevertheless, automatic analysis of February detected presence of \textit{O. fuscifacies} in a sample that was not detected by manual labeling initially. Close observation of the sample in the position that the algorithm detected the call, enabled the researcher to identify and label that sample correctly in the case in which only one call existed. 

It is observed that the call detector performed well in terms of accuracy and precision while maintaining high specificity. In other words, the detector exhibited no false positives in the labeled set (false species presence) and did not confound between detected species. On the other hand, recall of 0.875 signifies that a species present in the recording was not detected. This behavior was mostly due to species calling once during the sample time and pose a limitation that can be solved by increasing the sampling time or using a species specific approach setting a lower threshold for the desired species. These results prove that the proposed learner is able to generalize to real world audio that has not been used for training and validation, and was recorded with different equipment.

%A total of 37326 segments were detected by the system with 28\% of detections classified as unknown and the remaining 72\% as one of the 10 available frog species models. The results of the automatic analysis are plotted for the 3 species with the highest number of detections per pond in Figure~\ref{fig:pond1} and Figure~\ref{fig:pond2}.%
Overall, more detections occurred in Pond 1 than in Pond 2 suggesting higher acoustic activity in Pond 1 during the duration of the study. In Pond 1, the highest number of detections belonged to \textit{B. cinerascens} accounting for 58\% of detections followed by \textit{D. bifurcus} with 23\% and \textit{L. discodactylus} with 17\%. The remaining 2\% belonged to \textit{O. fuscifacies} and \textit{B. lanciformis}.  In contrast, Pond 2 shown mostly detections of \textit{D. bifurcus} and \textit{E. petersi}.  

\begin{figure}[!htb]

    \centering
   % \begin{subfigure}[!t]{0.48\textwidth}
   %     \centering
        \includegraphics[width=0.75\textwidth]{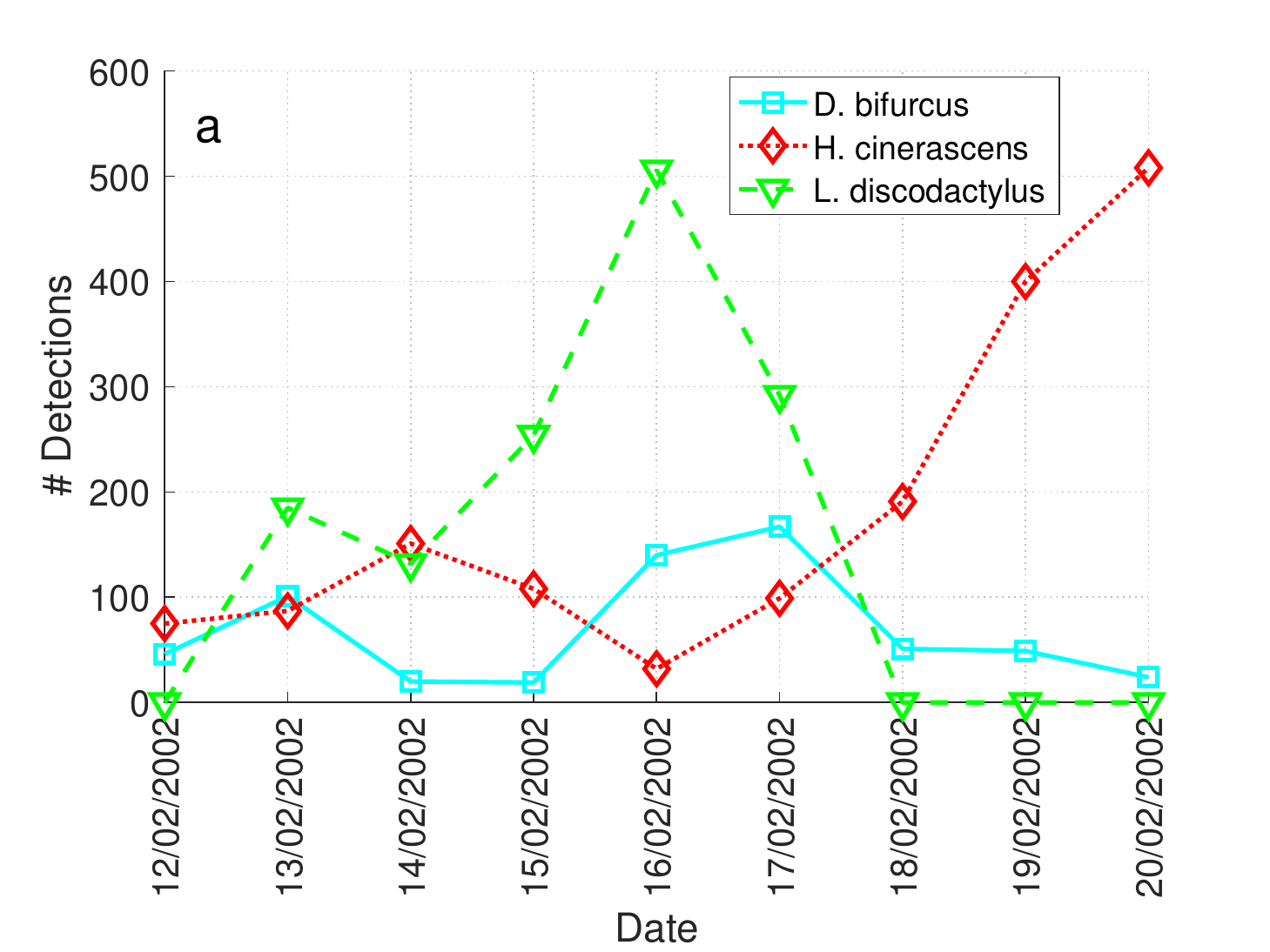}
        %\caption{Lorem ipsum}
   % \end{subfigure}%
 %   ~ 
 %   \begin{subfigure}[!t]{0.48\textwidth}
 %       \centering
 %       \includegraphics[width=1\textwidth]{pond1_04.eps}
 
 %   \end{subfigure}
 %   ~
 %   \begin{subfigure}[!t]{0.48\textwidth}
 %       \centering
 %       \includegraphics[width=1\textwidth]{pond1_07.eps}
 %    \end{subfigure}
 %   ~
 %   \begin{subfigure}[!t]{0.48\textwidth}
       
 %      \centering
 %       \includegraphics[width=1\textwidth]{pond1_08.eps}
 %    \end{subfigure}
 %        ~
 %   \begin{subfigure}[t]{0.48\textwidth}
       
  %     \centering
  %      \includegraphics[width=1\textwidth]{pond1_09.eps}
  %   \end{subfigure}

    \caption{Number of frog call detections on Pond 1 at YNP-PUCE station during the daily 10-minute recordings performed during a) February}%, b)April, c)July, d)August and e)September of 2002.}
         \label{fig:pond1}
\end{figure}

Figure \ref{fig:pond1} presents the number of detections of the three species which called the most during the sampling of February in Pond 1. A researcher can gain insights about the reproductive activity of those species with longer and planned acoustic samples. For instance, the circadian reproductive activity, and probably abundance might be extracted. However, it is still not clear how to extrapolate the number of males calling to the actual population including females and juveniles with the usage of the proposed approach in YNP.

\begin{figure}[!htb]
    \centering
        \makebox[\textwidth][c]{\includegraphics[width=1.25\textwidth]
        {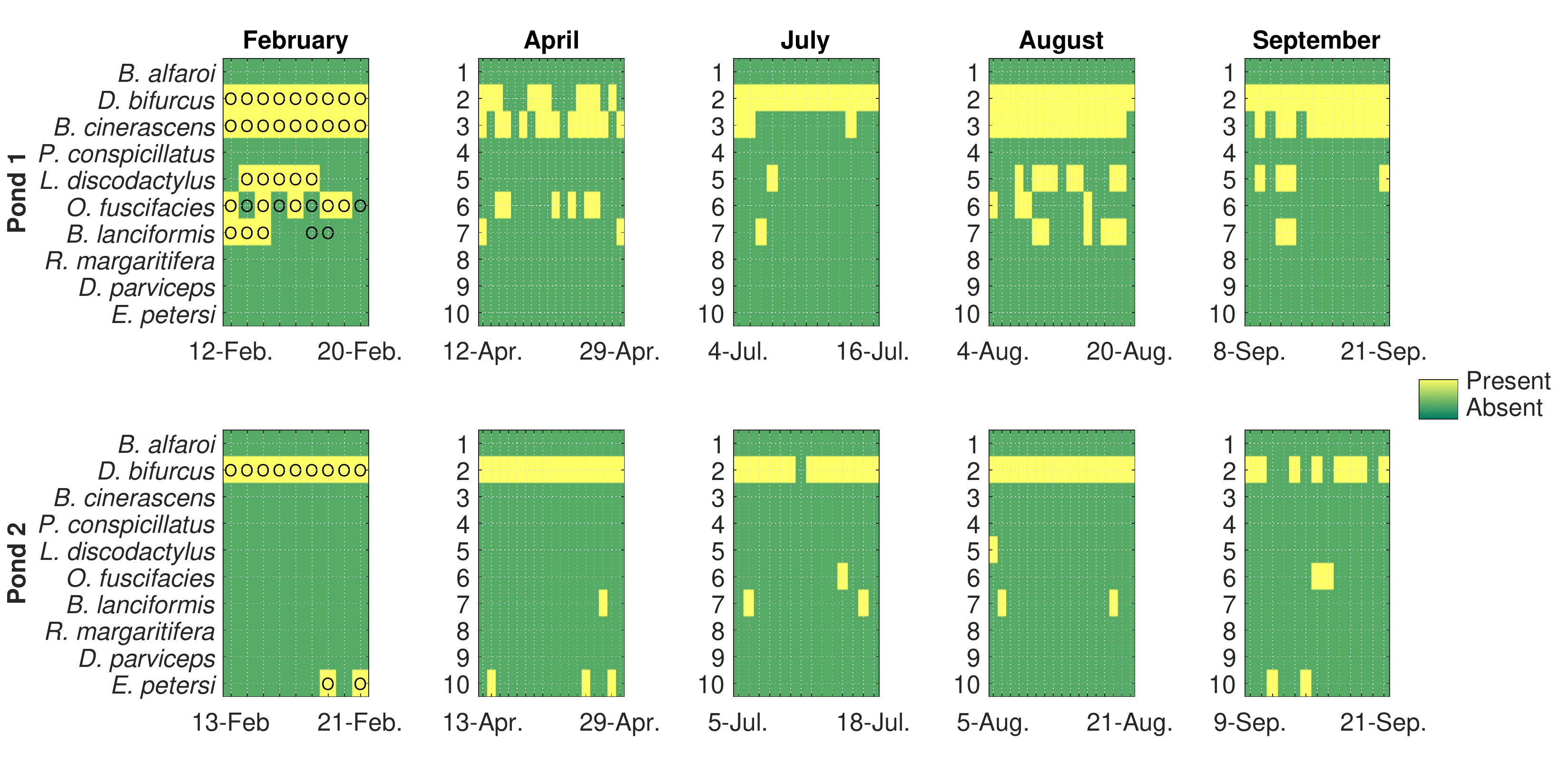}}
        \caption{Results of automatic 10 species presence-absence estimation in the available corpus. Species detected manually are shown by a circle.}
    \label{fig:summary}
\end{figure}

Finally, results in Figure \ref{fig:summary} gives a clue of the possibility to study seasonality and species richness in long-term sampling using acoustic methods in tropical frog communities. This is an interesting topic currently explored by ecologists\cite{diechman}.

\section{Discussion}
\label{section:Discussion}

%We explored the performance of human speaker recognition technology and adapted it to the problem of call-independent frog species recognition in a tropical forest. The acoustic feature set was modified to reflect the properties of frog calls and open-set recognition was applied to account for the high sound diversity commonly found in tropical soundscapes. The system introduces the unknown class option that allows researchers to focus on a smaller set of sounds to perform manual analysis rather than the all dataset.  

Monitoring animal sound in the tropical rainforest using automatic approaches is known to be challenging problem because of the high amount of noise present and variable conditions\cite{riede_monitoring_1993,XIE201713}. Currently, many algorithms have been proposed that allow researchers to study audio in search of calls of birds\cite{Ulloa2016}, insects\cite{Potamitis2014c}, odontocetes\cite{roch_gaussian_2007}, and frogs\cite{bedoya_automatic_2014,aboudan_acoustic_2013,XIE201713}. However, more research is needed to asses their suitability to study audio recordings made in frog communities with high biological diversity such as Yasun\'i National Park of Ecuador. An automatic approach to estimate presence-absence in long-term audio recorded on the site with local taxa is necessary to help researchers gain understanding of the dynamics and ecology of local frogs.

In this study, we applied frog call recognition with verification stage to study real-world audio recordings. Frog call classification have been attempted in selected calls with high SNR in previous studies \cite{bedoya_automatic_2014,brandes_feature_2008} showing that calls with low SNR were misclassified due to interference and noise. We found that it is not necessary to detect all the frog calls in the audio sample to achieve the goal of 10-species presence-absence estimation. As long as one call is detected, it is enough for species presence, which is highly probable in ponds where frogs call repeatedly to attract mates. Threshold setting then becomes an important step to tune the detector and set a desired operating point which minimize false alarms. Nonetheless, for frogs that called once during the sample, a limitation of this approach was observed which was also the case during human annotation by AET using headphones. For instance, calls of \textit{O. fuscifacies} and \textit{B. lanciformis} that occurred once or twice during the sample were not heard, and in consequence not labeled the first time. Observing the results of the algorithm in those samples,  which shown detections of those species, enabled the researcher to go back to the recording and verify that they were really present. In that context, the algorithm already was helpful to complement human performance and save time for labeling acoustic samples, which is an important task to prepare ground truth for developing Machine Learning algorithms.  

Even though previous studies suggested that MFCC coefficients are not suited for animal call recognition\cite{Towsey2012,roch_gaussian_2007}, they used MFCCs for voice recognition without modification. We found that a filter-bank modification based on the spectral content of frog-calls enabled the resulting cepstral features to improve classification performance using GMMs. This is an important result since it suggests that hand-crafted cepstral coefficients perform better when focused on the spectral characteristics of the taxa of interest. Despite efforts to develop a one-suits-all system, generally algorithms that perform well in some situations tend to do poorly in other datasets as stated by the \textit{no free lunch} theorem. We applied our approach to two distinct audio samples recorded at different ponds within YNP with consistent results, suggesting a good generalization capability. In addition, the audio used for training the GMMs was captured with digital equipment whereas the audio used for testing was recorded using an analog cassette recorder with different microphones. This result is important since suggests the possibility of studying audio coming from different sources and equipment, which is normally the case in the field where multiple people record calls in different timestamps.

Front-end processing is very important to obtain the results shown in this study. In \cite{XIE2016627}, Xie et al. proposed a frog acoustic activity detector in order to focus their classifier only in frog calls. We aimed to keep front-end segmentation as simple as possible to allow verification after classification to remove non-frog call audio. Verification was able to reject noise coming from malfunctioning cable, human voice, unknown species, calls overlap, cellphone noise, etc. The STE segmentation approach applied proved good for situations where frogs call intermittently with at least 10 ms of inactivity between calls. In species like \textit{D. bifurcus} which call in choruses in a non-stop way a limitation was identified. Since this is a variable size segment approach focused on a band of interest, it could be improved if multi-frequency segmentation is applied as suggested in \cite{inproceedings}, and classification and verification applied to each resulting sequence and adding the species detected. Finally, recent advances in end-to-end convolutional neural networks CNNs and recurrent neural networks (RNN), that study all the audio without prior segmentation could provide a way to remove front-end processing altogether. Nonetheless, the  processing power needed for that approach might be prohibitive when using desktop computers, and require cloud processing that is expensive. Research in Deep Neural Networks applications is advancing fast and we expect that using that approach important results will be obtained in the future. Therefore, we open the data-set to the research community \cite{dataset} and provide a baseline for comparison.

\section{Conclusion}
\label{section:Conclusion}

The proposed approach proved a helpful tool in estimating presence-absence of frog species in pond recordings made in the wilderness of YNP in Ecuador. Several hundred hours of unidentified acoustic material still exist at QCAZ-PUCE archive, and the application of automatic analysis could save researchers' time with metadata generation that can be verified in a fraction of the time that takes to listen each recording one-by-one. Fast audio appraisal and inventory generation without the need of specialists can be performed with acceptable level of performance by using metadata from presence-absence estimation. In systematic Ecoacoustic recordings used in wildlife monitoring, summing the results of multiple learners trained with specific taxa cohorts might provide a way to estimate biodiversity and study the composition of the soundscape. However, it is difficult to deeply asses those applications at this stage without more acoustic data available.  %Finally, for records spanning days of audio at a time, dividing the audio into smaller segments and summing up the results per segment could be a solution.    

Machine Learning aid in the automatic evaluation of frog communities in wildlife recordings introduces a potent technology that is complementary to existing survey techniques used currently by researchers in the wild. Our team is exploring diversity indexes estimation based on applying the proposed approach to 24-hour-long recordings made in the Mindo region in which critically endangered frog species call in a different setting with more silence between calls, which makes it simpler for front-end segmentation to extract calls from the background; thus, simplifying the digital signal processing pipeline.   %Its applications range from assessing frog communities composition, estimation of active male populations in a study zone, characterization and inventory of unidentified archived recordings and for creating automatic meta-tags used in automatic search. However, more research in needed to fully support those applications. 

Finally, automatic analysis of audio records of frog communities might be useful for researchers studying environmental changes since frog presence is related to the health of the ecosystem, and their disappearance provides clues of contamination or climate change effects that could be helpful in developing sustainable solutions.  Important applications in wildlife surveillance are envisioned that could be enhanced by wireless acoustic sensors networks in the wilderness.

\begin{figure}[!htb]
    \centering
        \makebox[\textwidth][c]{\includegraphics[width=1.66\textwidth,angle =90]
        {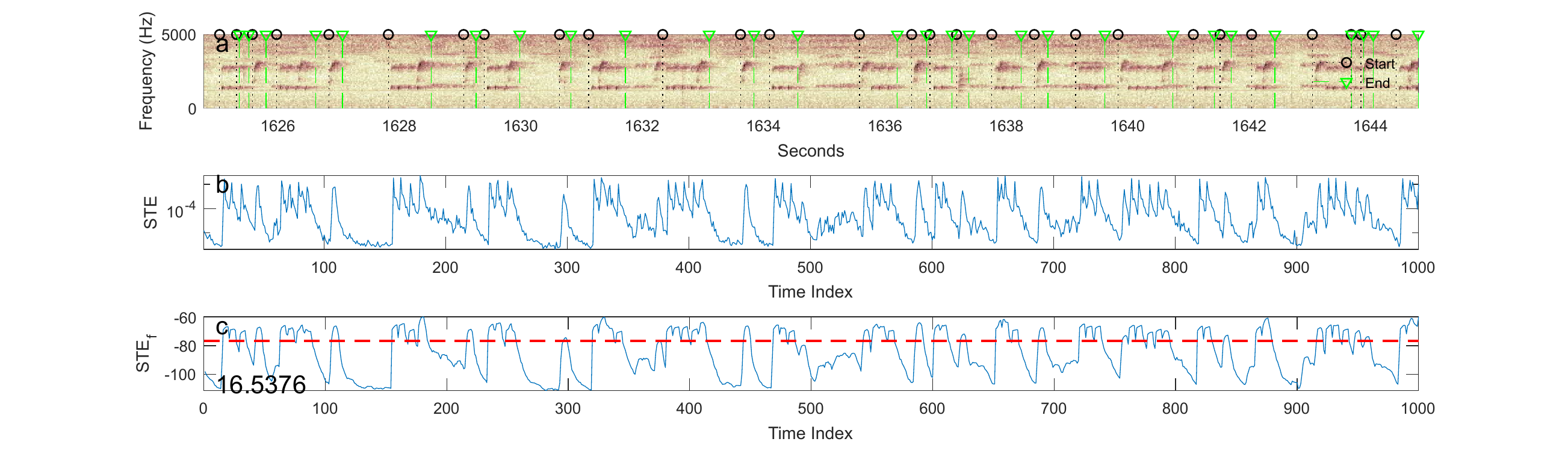}}
        \caption{30 seconds sample of Pond 1 recording. a. Segmented spectrogram, b. STE, c. Threshold}
    \label{fig:longrecordexample}
\end{figure}

% * <andrewsrocks@yahoo.com> 2018-07-16T21:06:58.832Z:
% 
% The segments classified as unknown by the system might contain species that are  present in the pond but whose calls are not  
% 
% ^.

\section*{Acknowledgments}

The study was supported by Pontificia Universidad Cat\'olica del Ecuador research projects 2015. Project L13304: "Dise\~no de un algoritmo de procesamiento de audio para un sistema que permita la automatizaci\'on del inventario, caracterizaci\'on y el monitoreo de poblaciones de las ranas del Ecuador, caso de estudio Parque Nacional Yasun\'i". We would like to thank Santiago Ron for providing access to the database of frog calls recorded in Yasun\'i National Park and for his valuable advice. To the staff at Estaci\'on Cient\'ifica Yasun\'i for their cooperation during the execution of field recordings. To Jean Camino, Franco Cisneros and Eduardo Silva for their important contribution in manual labeling of the frog calls used in training and development. To Daniela Pareja and Daniel Rivadeneira for their help during field work recording frog calls at YNP trails. To Samael Padilla for kindly providing access to the data gathered during his thesis work. To Paloma Lima for her help in early stages of the manuscript. 

\section*{References}

\bibliographystyle{elsarticle-num}
\bibliography{bibfile}

\end{document}